\documentclass[aps,prb,twocolumn,flushbottom,longbibliography,superscriptaddress]{revtex4-1}
\usepackage{graphicx}
\usepackage{dcolumn}
\usepackage{bm}
\usepackage{amssymb}
\usepackage{amsmath}
\usepackage{mathtools}
\usepackage{color}

\begin{document}

\title{Path integral Monte Carlo simulation of global and local superfluidity in liquid $^{4}$He reservoirs separated by nanoscale apertures}

\author{Tyler Volkoff}
\email{volkoff@snu.ac.kr}
\affiliation{Berkeley Quantum Information and Computation Center and Department of Chemistry, UC Berkeley, Berkeley, CA 94720, U.S.A.}
\affiliation{Department of Physics, Konkuk University, Seoul 05029, Korea}
\author{Yongkyung Kwon}
\email{ykwon@konkuk.ac.kr}
\affiliation{Department of Physics, Konkuk University, Seoul 05029, Korea}
\author{K. Birgitta Whaley}
\affiliation{Berkeley Quantum Information and Computation Center and Department of Chemistry, UC Berkeley, Berkeley, CA 94720, U.S.A.}
\date{\today}

\begin{abstract}
We present a path integral Monte Carlo study of the global superfluid fraction and local superfluid density in cylindrically-symmetric reservoirs of liquid $^{4}$He separated by nanoaperture arrays. 
The superfluid response to both translations along the axis of symmetry (longitudinal response) and rotations about the cylinder axis (transverse response) are computed, together with radial and axial density distributions that reveal the microscopic inhomogeneity arising from the combined effects of the confining external potential and the $^4$He-$^4$He interatomic potentials.  We make a microscopic determination of the length-scale of decay of superfluidity at the radial boundaries of the system by analyzing the local superfluid density distribution to extract a displacement length that quantifies the superfluid mass displacement away from the boundary. 
We find that the longitudinal superfluid response is reduced in reservoirs separated by a septum containing sufficiently small apertures compared to a cylinder with no intervening aperture array, for all temperatures below $T_{\lambda}$.
For a single aperture in the septum, a significant drop in the longitudinal superfluid response is seen when the aperture diameter is made smaller than twice the empirical temperature-dependent $^4$He healing length, consistent with the formation of a weak link between the reservoirs.
Increasing the diameter of a single aperture or the number of apertures in the array results in an increase 
of the superfluid density toward the expected bulk value. 
\end{abstract}

\maketitle

\section{\label{sec:intro}Introduction}
Unlike the case of normal-metal weak links between superconductors or weak links between reservoirs of liquid $^{3}$He, the construction of
nanoscale superfluid weak links with engineered geometry supporting a robust Josephson current between reservoirs of liquid $^{4}$He remains a considerable engineering
challenge. \cite{packardnew} The relative difficulty 
in $^4$He is due to the much smaller value, by orders of magnitude, of the healing length $\xi$ in liquid $^{4}$He relative to that in BCS-type systems when both are
deep in their respective condensed phases.  For example, the measured value of $\xi(T=0)$ is on the order of 1.0 $\mu$m for a type-II BCS superconductor \cite{nakahara} and 64 nm for $^{3}$He (at 0 bar of pressure), compared with only 0.3 nm for $^{4}$He. \cite{packard3}
A weak link, which generically consists of two condensed fluids separated by a junction consisting of noncondensed matter, e.g., the same fluid in a noncondensed phase,
requires that the mass supercurrent through the connection in response to an external field (e.g., 
mechanical driving in neutral superfluids, voltage bias in superconductors) is much smaller than the response of the bulk
supercurrent. 
If two reservoirs of liquid helium 
are separated by an array of nanoapertures with an average cross-sectional diameter 10 nm, experimental studies have shown that such an array supports a weak link only at temperatures $T$ very close to the lambda transition, $T_{\lambda} \approx 2.17$K, specifically, at temperatures $T$ such that $T_{\lambda} - T \lesssim 0.05$ mK. \cite{packard}
This is consistent with filling of the apertures by normal fluid as the lambda transition is approached from below.

Engineering of aperture arrays with apertures having much smaller diameter (e.g., $\mathcal{O}(10\, \textrm{\AA})$) would allow experimenters to probe phenomena associated with $^{4}$He weak link formation, e.g., Josephson oscillations deep in the superfluid phase, and strongly-interacting quasi-2D and quasi-1D liquid $^{4}$He in precisely controlled geometries. For example, it has been shown theoretically using both mean-field methods and the classical pendulum analogy for the d.c. Josephson equation 
that the presence of multiple apertures in an array can lead to reduction of decoherence in the macroscopic phase
differences across the array. \cite{pekker,chui} In addition, quantum field theoretical analysis of tunnel-coupled reservoirs of interacting bosons predicts a linear scaling of the amplitude of Josephson oscillations with the number of apertures in an array. \cite{volkoff} An understanding of superfluidity in engineered nanoaperture arrays is central to exploring and generalizing these predictions. Nanoaperture arrays separating reservoirs of liquid $^{4}$He are 
also an essential component of proposals for  highly sensitive rotation sensing devices based on matter wave Sagnac interferometry. \cite{packsat,sato, joshi}

By revealing the distribution of superfluidity  
for liquid $^{4}$He confined in nanoscale and atomic scale potentials, numerical simulation of confined liquid $^{4}$He below the lambda transition temperature aids the design and implementation of aperture arrays for superfluidity experiments. In this work, we use a path-integral Monte Carlo (PIMC) algorithm to compute the global superfluid response and local superfluid density of a $^{4}$He reservoir consisting of $N=35-123$ $^{4}$He atoms confined by a cylindrically symmetric tube potential with diameter  
$16-20\,$\AA~and length  
$18-24\,$~\AA. We then introduce a septum containing 
one or more nanoscale apertures, thereby breaking the translational invariance of the potential, and we use PIMC to calculate the global and local superfluid response in the resulting system of connected reservoirs. We present a detailed analysis of the longitudinal and transverse superfluid responses for a single aperture as a function of temperature, aperture size, and location, and then we study  aperture arrays with up to $N_a = 5$ apertures under conditions in which the single aperture shows behavior consistent with a weak link.

In addition to analyzing the superfluid responses of these  
reservoirs connected by nanoscale apertures, we undertake a microscopic analysis of the length scale characterizing decay of superfluid density at the radial boundary of the cylindrical reservoirs.  This length scale might be considered to coincide with the boundary-induced decay length for superfluid density that appears in the Ginzburg-Pitaevskii (GP) theory of superfluidity \cite{pitaevskii}, 
which is one of a number of measures of the length scale over which the superfluid response drops to zero at a boundary. Any such measure is commonly referred to as a ``healing length".  Experimental studies on bulk $^4$He generally measure the healing of a superfluid with the empirical temperature dependent expression first determined  
from measurements of superfluid density in $^{4}$He films flowing through a slit 
of $\sim3900$~\AA~ spacing formed by two concentric cylinders. \cite{reppy} This empirical healing length $\xi$ is given by 
\begin{equation}\label{eqn:healing} 
\xi(t)=0.34 \textrm{ nm}/ t^{0.67} 
\end{equation} 
with $t=(1-T/T_{\lambda})$ the reduced temperature. \cite{reppy,packard,goldner} With some modifications to approximately reproduce the correct critical exponents for the empirical healing length and the superfluid density  \cite{ginzburgsobyanin,slyusarev}, GP theory may be applied to such bulk systems when the fluid density is assumed to be homogeneous and the superfluid is assumed to decay over macroscopic length scales.   However, a GP approach is not applicable when the fluid density shows atomic scale microscopic variations.  Indeed, even in the bulk, the healing length of a spatially inhomogeneous superfluid is not uniquely defined \cite{donnelly}, creating a challenge for measurement and quantification of the decay of superfluid density in nanoscale confined systems.

In this work, we show that the temperature $T$ at which the empirical $\xi(T)$ equals the aperture radius 
is qualitatively a predictor of the formation of a weak link between reservoirs of liquid $^{4}$He separated by a nanoscale aperture.  Since $\xi(T)$ is not defined microscopically and GP theory does not apply to the atomic scale superfluid density oscillations observed in these nanoscale confined systems, we quantify the ``healing" of the superfluid instead by a local displacement length that measures the superfluid mass displacement by the boundary. \cite{hills,volkoff}   In particular, the displacement length that we introduce can be computed directly from PIMC data and is unequivocally defined for non-translationally invariant, cylindrically symmetric confined superfluids. We further show that this local displacement length can be applied to generate a ``healing surface,'' which is a useful notion for visualizing the displacement of superfluidity from high-potential regions in generic symmetric or irregularly-shaped systems.

A brief outline of the paper is as follows: in Section \ref{sec:bareresults}, we analyze the radial and axial atomic density distribution, the temperature dependence of the superfluid response, and the displacement length at the system boundary in 
longitudinally translation-invariant cylinders.
This allows a comparison with previous 
calculations on the radial distribution of superfluidity of liquid $^{4}$He confined by nanopores. \cite{maestro13}
In Section \ref{sec:apertureresults}, we then present our results for the global superfluid fraction and local distribution of superfluidity of liquid $^{4}$He in a tube that is partitioned by a septum containing nanoscale apertures into two communicating reservoirs.  We focus first on a single aperture, characterizing the effect of this in detail, and then we 
present results for nanoaperture arrays containing up to five apertures of cross-sectional diameter $5\,$\AA~that are arranged in various spatial configurations in the plane of the bipartition. 
Section \ref{sec:conclusion} summarizes and discusses the implications for analysis and design of experiments with helium superfluid flow through nanoscale aperture arrays. 

The numerical simulations in this work are carried out at specified temperatures ($0.25 \textrm{ K} \le T \le 2.0 \textrm{ K}$) and 
calculated pressures ($\sim 3.5$ bar, calculated with the estimator of Ref.\onlinecite{ceperley95}) that lie well within the bulk He II superfluid phase. For all simulations, a periodic boundary condition is imposed along the longitudinal direction,  i.e., along the tube axis, to minimize finite-size effects and allow for simulation of estimators of certain physical observables that depend on the longitudinal winding number of the  
imaginary time paths of the $^{4}$He atoms in the PIMC 
calculations.

\section{\label{sec:bareresults}PIMC calculations for translation-invariant, cylindrically symmetric potential}

For our PIMC study of $^4$He atoms contained in a nanoscale tube, we define the following potential:
\begin{equation}
V_{\textrm{{\footnotesize tube}}}(R,\phi,z)=\frac{V_0}{2}\Big[1+\tanh\Big(\frac{R-R_t}{\sigma_R}\Big)\Big] ,
\label{eq:tube}
\end{equation}
where the cylindrical coordinates $R$, $\phi$, and $z$ represent the distance from the tube center, the azimuthal angle, 
and the coordinate along the tube axis, respectively, and $R_{t}$ is the radius defining the onset of a large ``wall'' potential. In terms of Cartesian coordinates $(x,y,z)$, $R=\sqrt{ x^{2} + y^{2}}$ and $\phi = \tan^{-1}\left( {y\over x}\right)$.
In Eq.(\ref{eq:tube}), $V_0$ (potential strength) and $\sigma_R$ (steepness of potential) are taken to be $150$~K and $0.25$~\AA, respectively.
$V_{\textrm{{\footnotesize tube}}}(R,\phi,z)$ is independent of $z$ and so it is invariant under translations along the tube axis.
For the $^4$He-$^4$He interaction, we use a well-known Aziz potential. \cite{aziz92}
In the path integral representation, the thermal density matrix at a low temperature $T$ 
is expressed as a convolution of $M$ high-temperature density matrices with 
an imaginary time step $\tau = (M k_{B} T)^{-1}$.  
In the high-temperature density matrix 
the $^4$He-$^4$He potentials are incorporated with the pair-product form of the exact two-body density matrices
while the external potential defined by Eq.(\ref{eq:tube}) is analyzed within the primitive approximation. \cite{ceperley95}
We use a time step of $\tau^{-1}/k_{B} = 40$ K and periodic boundary conditions are imposed in the $z$ direction to minimize finite size effects.

\begin{figure}
\begin{center}
\includegraphics[scale=.35]{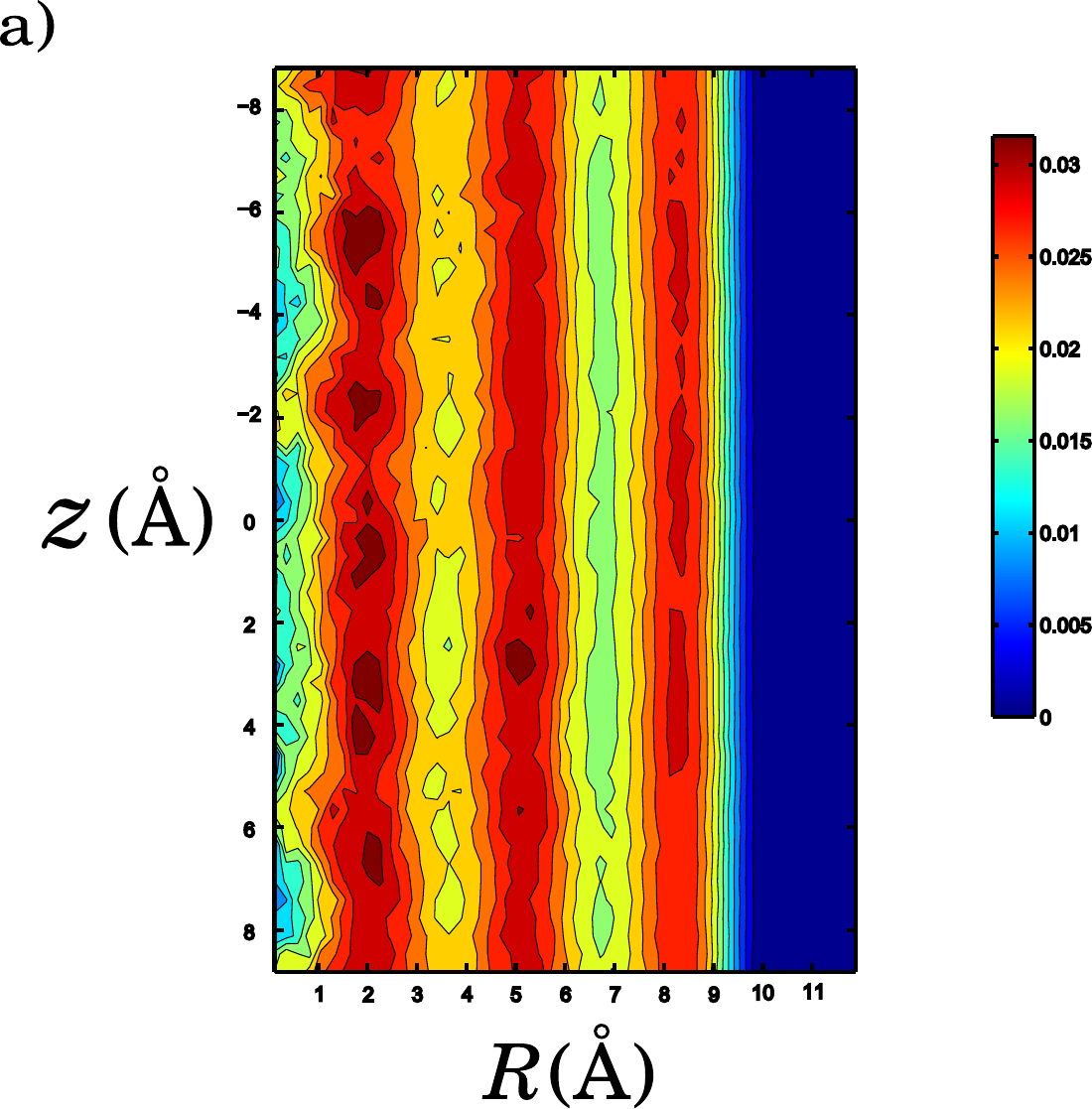}
\includegraphics[scale=.38]{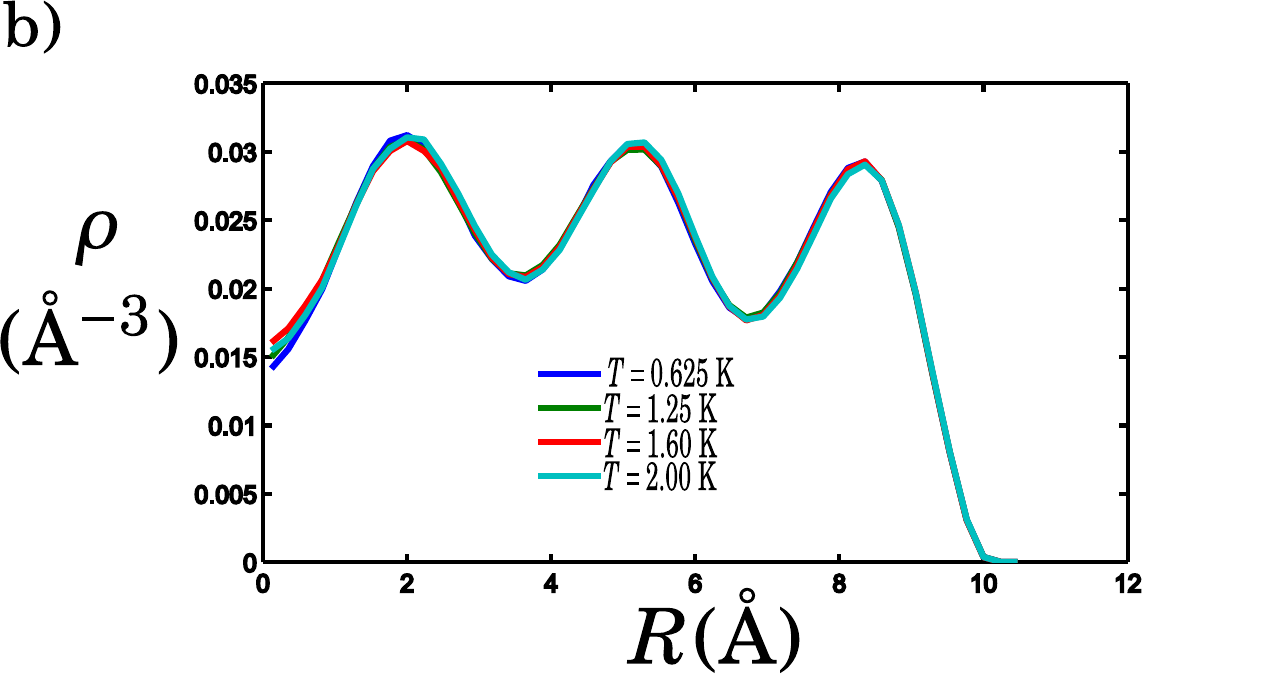}%
\caption{\label{fig:tube_den} a) 2-D density distribution of $^4$He atoms at $T=1.25 K$ contained inside a tube of radius $R_{t}=10$~\AA~and length $L = 18$~\AA, with periodic boundary conditions along $z$.
The atomic density distribution $\rho(R,\phi,z)$ is averaged over the azimuthal angle $\phi$ to give $\rho(R,z)$ in units of~\AA$^{-3}$  (red: high density, blue: low density).
b) One-dimensional density distributions in (units of~\AA$^{-3}$) computed as a function of $R$ for a range of temperatures below the bulk $T_{\lambda}\approx 2.17$~K.
}
\end{center}
\end{figure}

We first computed the density distributions of $N=123$ $^4$He atoms contained by the potential of 
Eq.~(\ref{eq:tube}) with the tube radius $R_t$ set to be 10 \AA~ and length $L= 18$ \AA. 
Note that the lowest value of the tube potential of Eq.~(\ref{eq:tube}) occurs at the center of the tube ($R=0$)
and that $V_{\textrm{{\footnotesize tube}}}$ increases monotonically as $R$ increases.  
Figure~\ref{fig:tube_den} a) shows a contour plot of the density distribution at $T=1.25 K$
averaged over the azimuthal angle $\phi$.  One can observe 
a layering structure around the tube axis ($R=0$).
This layering is 
due to the 
interplay between the $^{4}$He-$^{4}$He interparticle interaction
and the atomic confinement due to the nanoscale confining potential of Eq.~(\ref{eq:tube}).  
It is to be contrasted with the layering observed 
in PIMC calculations 
for $^4$He atoms inside an amorphous Si$_3$N$_4$ nanopore  
that included 
attractive van der Waals interactions 
of $^4$He with the pore wall. \cite{maestro13}
While that work also found layered structures for $^4$He inside the nanopore, the layering
there was primarily 
due to the interplay of the repulsive $^4$He-$^4$He interaction and the  
attractive component of the $^4$He-wall interaction. The latter provided attractive adsorption sites for $^4$He 
in the vicinity of the wall that caused the two outermost $^4$He layers to be solidified 
without making any contribution to superfluidity at low temperatures.
The fact that our calculations 
do not assume an attractive van der Waals interaction between the $^{4}$He atoms and an atomically-structured  wall implies that the layering of the $^{4}$He density shown is then due solely to the interplay of the $^{4}$He-$^{4}$He interaction with the external confining potential. As a consequence, no solidified layers are observed. The absence of solidified layers allows an unambiguous characterization of the superfluid healing behavior at the boundary of a nanoscale container (see Section \ref{sec:spatdist}).
Figure~\ref{fig:tube_den} b) shows the one-dimensional density distributions computed as a function of radius $R$,
for several temperatures between 0.625~K and 2~K. It is evident that there is no thermal effect on the $^4$He density distribution at temperatures below 2~K.

The global superfluid response to translations of the system along the tube axis, i.e., the $z$-axis,     
may be computed by using the following winding number estimator for the global superfluid fraction \cite{ceperley95}:
\begin{equation}
\left( \frac{\rho_{s}}{\rho}\right)_{z}=\frac{m L^2 \langle W^{2}_{z} \rangle }{\hbar^2 \beta N} \, ,
\label{eq:windsf}
\end{equation}
where $m$, $L$, and $N$ are the bare mass of a $^4$He atom, the length of the tube, 
and the number of helium atoms inside the tube, respectively.
Here the winding number $W_{z}$ is defined by $W_{z}= 1/L \sum_{i=1}^N \sum_{k=1}^{M} (z_{i,k+1}-z_{i,k})$,
where $M$ is the number of time slices in the discrete path-integral representation, the sums are over particle index $i$ and imaginary-time slice index $k$, and $z_{i,k}$ is the projection of the 
single particle imaginary time configuration $\vec{r}_{i,k}$ onto the cylinder axis. Therefore, a nonzero average winding number indicates the onset of superfluidity.

\begin{figure}
\begin{center}
\includegraphics[scale=.4]{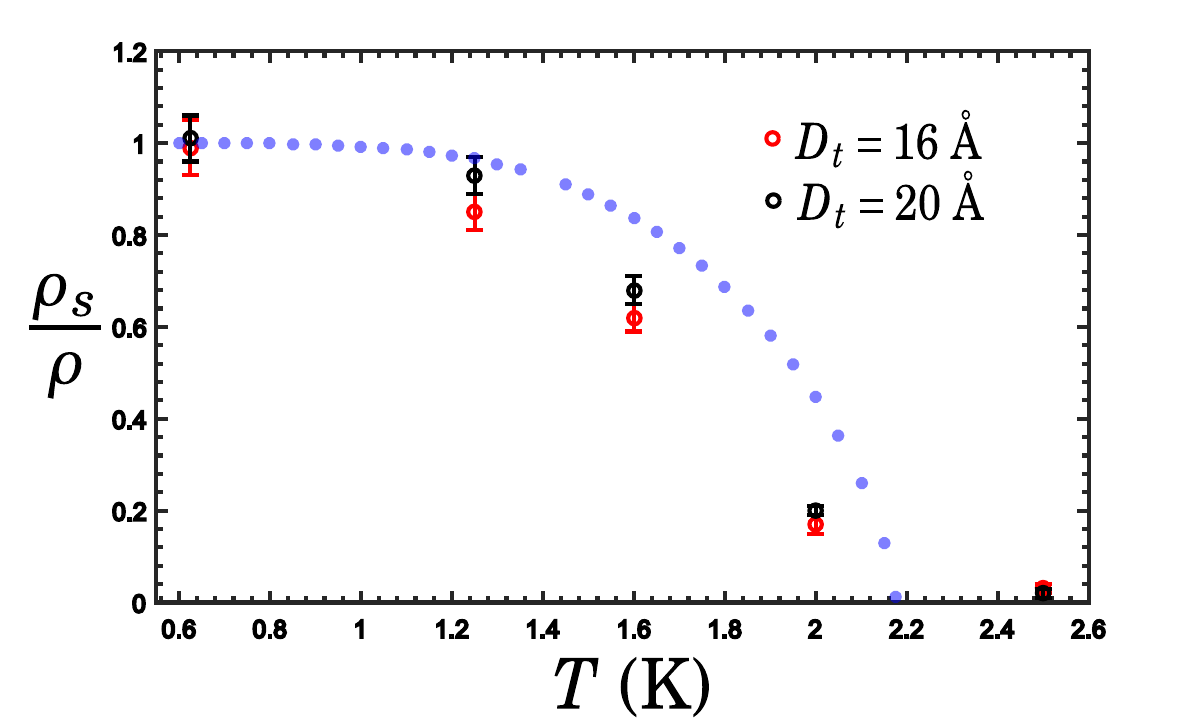}
\caption{\label{fig:tube_sf}Global superfluid fraction of liquid $^4$He contained 
inside tubes of diameter $D_{t} = 16$~\AA~(red circles) and $D_{t}=20$~\AA~(black circles)
as a function of temperature.
The calculations were made for $N=80$ atoms (red circles) or $N=123$ atoms (black circles) in a cylinder of length $L=18$~\AA, with periodic boundary conditions in $z$. Blue dots show recommended values for the global superfluid fraction of bulk liquid $^{4}$He below $T_{\lambda}$ at saturated vapor pressure.\cite{donnellydata}
}
\end{center}
\end{figure}

Figure~\ref{fig:tube_sf} shows the global superfluid fractions
of liquid $^4$He inside tubes with two different diameters ($D_t=2R_t$) as a function of temperature. Recommended values for the expected superfluid fraction in bulk liquid $^{4}$He at and below $T_{\lambda}$  are also shown for comparison.\cite{donnellydata}
For temperatures below 1 K, both $^4$He systems are seen to exhibit complete superfluid response, similar to bulk liquid $^4$He. 
In contrast, the PIMC calculations of Ref.~\onlinecite{maestro13} for $^4$He atoms in a Si$_{3}$N$_{4}$ nanopore 
show a saturated superfluid fraction of only $\rho_{s}/ \rho \sim 0.2$ for $T \lesssim 1$ K, as a result of the inert solid $^4$He layers adsorbed on the pore wall. 
From a microscopic perspective, the fact that our calculations show complete superfluid response for the present $^4$He-nanotube system in this temperature range
is due to the fact that the tube potential of Eq.~(\ref{eq:tube}) does not adsorb layers of solid $^{4}$He.  This may also be interpreted within a hydrodynamic perspective, where the lack of short-range van der Waals attraction between the $^{4}$He atoms and the wall means that a longitudinal translation of the wall does not result in entrainment of  
any part of the liquid helium-4 as a solid layer; therefore, the mass fraction of the liquid that exhibits a superfluid response is higher in our calculations relative to Ref.~\onlinecite{maestro13}.
Reduction of superfluid fraction also occurs in the first layer of liquid $^{4}$He adsorbed on a molecular dopant embedded in a liquid $^{4}$He nanocluster.\cite{kwonwhaleyprl} In general, an attractive van der Waals interaction of helium atoms with an atomically-structured container or with an embedded dopant pins the imaginary-time paths at the surface-liquid or dopant-liquid interface and thereby renders unlikely the acceptance (in the Metropolis algorithm employed in PIMC)  of a permutation move combining paths of two or more $^{4}$He atoms to create an extended path that makes a non-zero contribution to the total winding number in Eq.~(\ref{eq:windsf}).  This effectively removes helium density from the superfluid component, resulting in a molecular scale non-superfluid 
density as first described in Ref.\onlinecite{kwonwhaleyprl}. Calculations of the global superfluid fraction of nanoscale liquid $^{4}$He confined by parametrized external potentials modeling adsorption at a system boundary have also been performed in spherical \cite{boninsegni} and cylindrical \cite{glydepimc} geometries.

In Figure~\ref{fig:tube_sf}, the superfluid fraction is observed to decrease for $T \ge 
1$~K in both $D_{t}=16$~\AA~and $D_{t}=20$~\AA~tubes, similar to the decrease observed in bulk $^4$He in Ref.\onlinecite{ceperley95},
except with comparatively lower values 
and broader transition to zero value, as is typical of finite size systems.
The quantum statistical explanation of this decrease in superfluidity inside the nanotube is the same as in the bulk, namely that as the temperature is increased toward the lambda transition, fewer winding paths contribute to the bosonic partition function,
as the thermal de Broglie wavelength and hence the exchange probabilities of the $^4$He atoms decrease. Therefore, the value of the winding number estimator in Eq.(\ref{eq:windsf}) also decreases.

For the nanotube it is then interesting to analyze the temperature at which the largest depression of $\rho_{S}/ \rho$ occurs in terms of the relationship between the corresponding healing length and the tube diameter. We find that for both values of tube diameter studied here, the temperature marking the onset of a large depression of global superfluid fraction is remarkably similar to the temperature at which the empirical healing length $\xi(T)$ of the superfluid approaches the tube radius $R_{t}$. Specifically, the empirical formula 
Eq.~(\ref{eqn:healing}) yields the following two temperature/healing length combinations 
for which the healing lengths are 
equal to half the tube diameters employed in Fig.~\ref{fig:tube_sf} : 
$\xi(T=1.56~\textrm{K}) = 8$~\AA, and $\xi(T= 1.74~\textrm{K}) = 10$~\AA. Hence, for $T\gtrsim 1.5$ K, the superfluid density 
inside the tube does not reach its maximal 
possible value 
and so the global superfluidity is further reduced below the bulk value in this regime.

However, while  $\xi(T)$ gives a consistent prediction of the temperature range at which the superfluid fraction falls 
significantly below 1, this empirical estimate of healing of the superfluid is not clearly related to the Ginzburg-Pitaevskii notion of decay of superfluidity near a boundary.
Moreover, as noted already above, within its domain of applicability the Ginzburg-Pitaevskii theory of superfluidity predicts monotonic decay of the superfluid density from its maximum value to zero, within a temperature-dependent distance from a flat boundary. \cite{pitaevskii} In Section \ref{sec:spatdist} below, we show that the radial superfluid density in a nanoscale cylindrically symmetric system does not decay monotonically from the axis of symmetry. We then provide a general method for quantifying the decay of superfluidity at such a boundary that, unlike the Ginzburg-Pitaevskii theory which is only valid for length scales much greater than the atomic scale, is now valid on all length scales.  
\begin{figure*}
\begin{center}
\includegraphics[scale=.50]{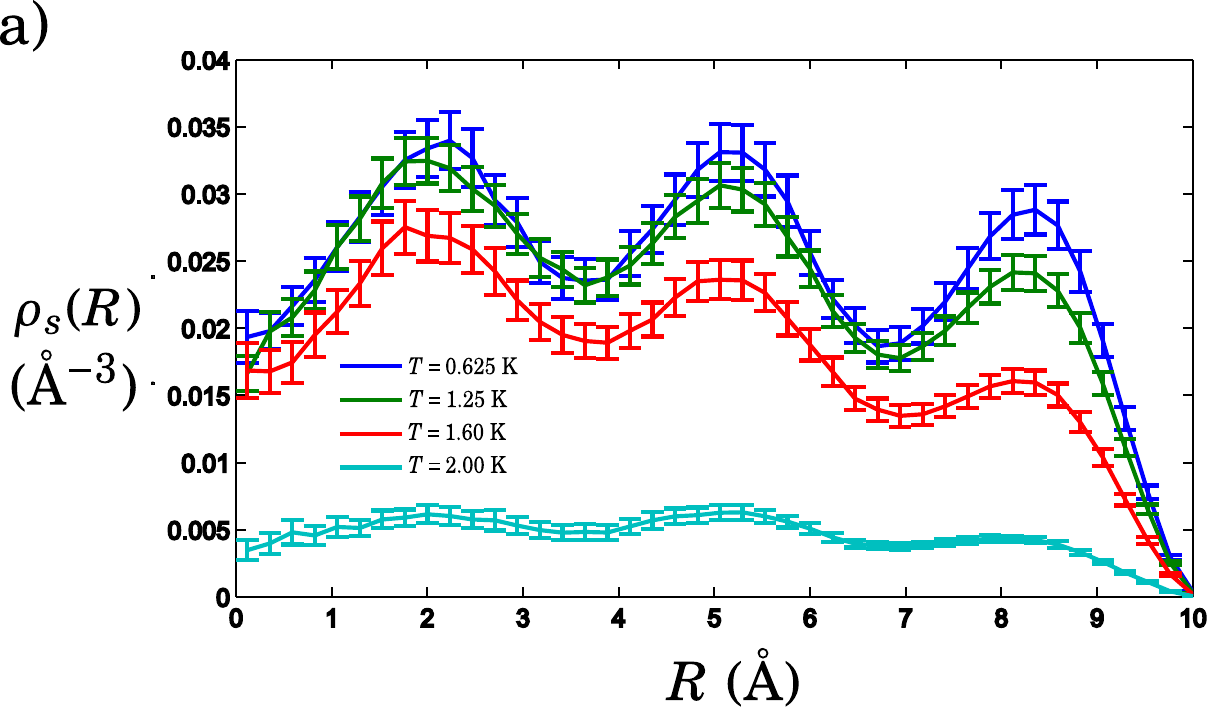}
\includegraphics[scale=.50]{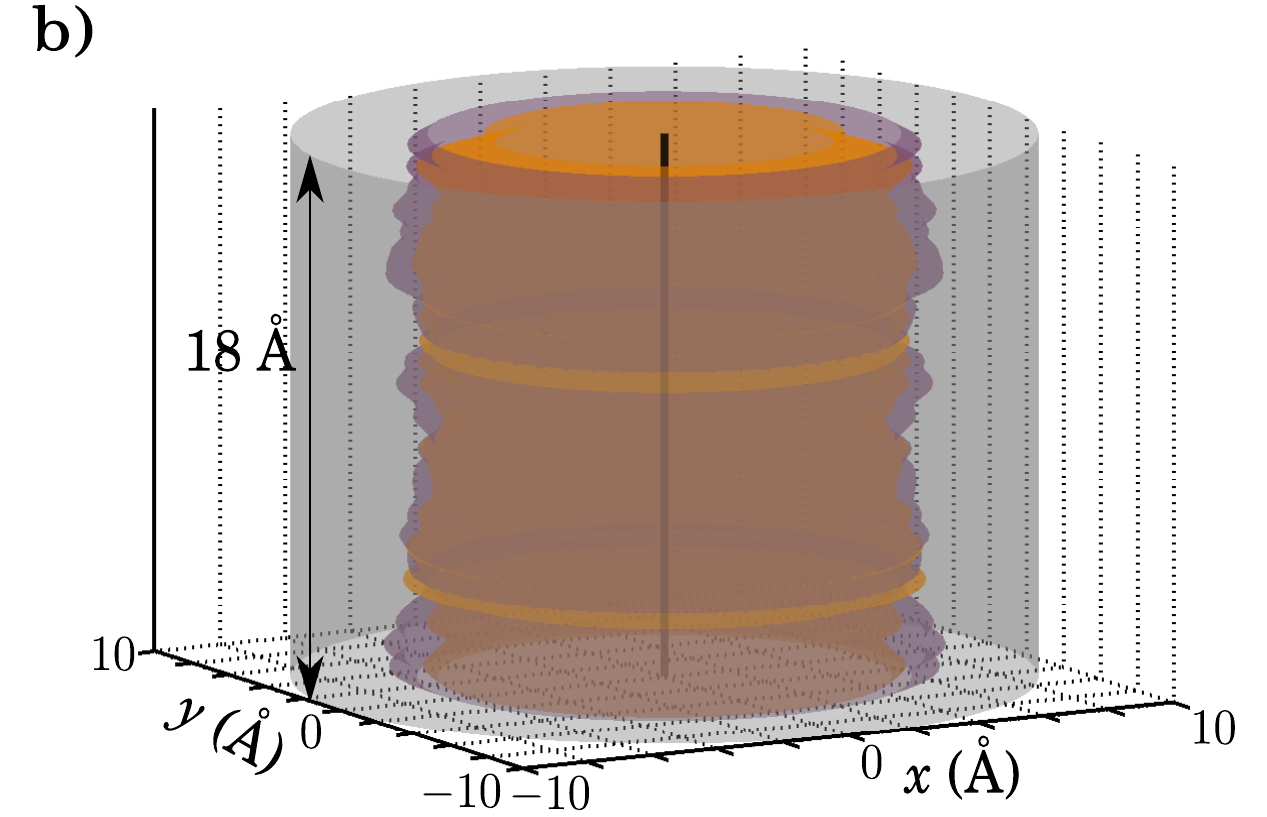}
\caption{a) One-dimensional radial superfluid density distributions in a cylindrically symmetric potential 
for four temperatures below bulk $T_{\lambda}$. Calculations were made for $N=123$ atoms in a tube of length $L =18$~\AA~and diameter $D_{t}=20$ \AA~with periodic boundary conditions imposed in the $z$ direction. 
The error bar at a given radial coordinate represents the sample standard deviation from the average superfluid density at that coordinate computed from statistically independent data blocks.
b) Healing surfaces for $T=0.625$ K (purple) 
and $T=2.00 K$ (orange) computed using a local version of the superfluid mass displacement length (see text). The outer cylinder 
with radius $10$~\AA~is a guide to the eye.
} 
\label{fig:tube_localsf}
\end{center}
\end{figure*}

\subsection{\label{sec:spatdist}Spatial distribution of superfluidity}
To analyze the spatial distribution of the superfluid density, we employ an estimator 
for the local longitudinal superfluid density, i.e., the superfluid response to translation in the $z$ direction, that is based on 
the following local decomposition of the winding number estimator in Eq.~(\ref{eq:windsf}):
\begin{equation}
\rho_s(\vec{r})_{z}=\frac{m L^2}{\hbar^2 \beta} \Big\langle \sum_{i =1}^{N_{w}} \sum_{k=1}^M 
\frac{W^{2}_{z}}{N_{w} M} \delta(\vec{r}-\vec{r}_{i,k}) \Big\rangle ,
\label{eq:local_sf}
\end{equation}
where $N_{w}$ is the number of $^4$He atoms comprising winding paths. This estimator of local superfluid density is similar to the local estimator of Khairallah and Ceperley \cite{khairallah} in the sense that all ``beads'' (represented by the coordinates $\vec{r}_{i,k})$
on the imaginary time ``polymers" constituting the winding paths are assumed to contribute equally to superfluidity.
Although the local estimator employed by Kulchytskyy {\it et~al.} \cite{maestro13} and the one used in this work
give the same proper value when integrated over space, that is, the global superfluid fraction
multiplied by the total number of $^4$He atoms, the two estimators for the local superfluidity
are based on different local decompositions of the winding number estimator 
(compare Eq.~(\ref{eq:local_sf}) of the present paper with Eq.~(3) of Ref.\onlinecite{maestro13}). 
In particular, Eq.~(\ref{eq:local_sf}) is locally positive semidefinite, which makes interpretation of regions 
of negative local superfluid density unnecessary, and it also exhibits less statistical noise, 
making the PIMC estimation more robust.

In Figure \ref{fig:tube_localsf}, we show the radial superfluid density, $\rho _{s}(R)$, computed using the local estimator of $\rho_{s}(\vec{r})_{z}$ 
in Eq.~(\ref{eq:local_sf}), and subsequently
averaging over the axial coordinate $z$ and angular coordinate $\phi$ of $\vec{r}$. The reported Monte Carlo error at each radial coordinate is the sample standard deviation from the local average value of the superfluid density. 
As the temperature increases, $\rho_{s}(R)$ is seen to decrease below the temperature-independent total density $\rho(R)$ in Fig. \ref{fig:tube_den} at all values of $R$. However, whereas the height of the peaks of $\rho(R)$ decreases only slightly as $R$ increases toward the boundary 
(see Fig.\ref{fig:tube_den} b)), the peaks of $\rho_{s}(R)$ in Fig.\ref{fig:tube_localsf} are 
noticeably reduced as $R$ increases, corresponding to a decrease in the superfluid fraction near the boundary.
The cause of this radial decrease of superfluid response is the fact that atoms distributed near the tube wall interact with fewer neighboring atoms,
so that a bosonic permutation move is less likely to be accepted near the wall than in central regions of the cylinder.  
This suggests that the decrease in superfluid fraction near the tube wall is a quantum statistical effect and is not caused by solidification of atomic layers at the wall due to an attractive potential.

For engineering nanoscale channels supporting superfluid helium flow, it is useful to quantify the length scale characterizing the decay of superfluid density at a region of large potential energy or a system boundary.  Because the empirical healing length $\xi(T)$ 
of Ref.~\onlinecite{reppy} is not directly computable from the local superfluid density distribution, one must identify a length scale that 
captures the same underlying physical features while at the same time satisfying three conditions: i) being extractable from the local superfluid density distribution, ii) increasing with increasing temperature, and iii) being applicable to non-translationally invariant potentials. There is no unique definition of such a length scale. Several local and global quantifiers of the displacement of superfluid density from the walls of a confined bosonic system were formulated from the local superfluid density distribution data and compared in Ref.\onlinecite{volkoffthesis}. In the present work, we adapt the notion of displacement length defined in Ref.~\onlinecite{hills} to define a local displacement length that is easily applied to the cylindrically symmetric systems considered here.

In a system of stationary liquid $^{4}$He occupying a half space $z\ge 0$ of $\mathbb{R}^{3}$ and satisfying the boundary conditions $\rho_{s}(z=0)=0$ and $\rho_{s}(\vec{r}) \rightarrow c=\, ${const.} as $z \rightarrow \infty $, the displacement length $d$ that quantifies the effective superfluid mass displacement at a planar interface $z=0$ is defined 
as
\begin{equation} 
\int_{\mathbb{R}^{3}}c\theta(z-d)dxdydz = \int_{\mathbb{R}^{3}}\rho_{s}(\vec{r})\theta(z)dxdydz.
\end{equation}
where $\theta(z) = 1$ ($z\ge 0$) and $\theta(z) = 0$ ($z<0$) is the step function.
The displacement length $d$ was shown in Ref.\onlinecite{hills} to scale near $T_{\lambda}$ as the reciprocal 
of the (roton) energy gap. To formulate a local version of the displacement length for cylindrically symmetric containers, we seek for each coordinate $z$ the distance $d(z)$ such that
\begin{eqnarray}
\left( \max_{0\le R \le R_{t}(z)}\rho_{s}(R,z)\right) \cdot \left( R_{t}(z) - d(z) \right) &=& \nonumber \\ \int_{0}^{R_{t}(z)}dR\; \rho_{s}(R,z) &{}&
\label{eqn:displength}
\end{eqnarray}
 is satisfied, where the tube radius at axis coordinate $z$ is defined by $R_{t}(z)$. In practice, the right hand side of Eq.(\ref{eqn:displength}) is evaluated by trapezoid rule integration of the spatially discrete $\rho_{s}(R,z)$ numerical data, which are in turn
 obtained by integrating the full numerical distribution $\rho_{s}(\vec{r})$ over $\phi$. $\rho_{s}(R,z)$ is defined to be the mean longitudinal superfluid density at the coordinate $(R,z)$.  In the present work, we consider only the simplest case of $R_{t}(z)=R_{t}$ for all $z$.
 
In the 
more general context of liquid $^{4}$He in a compact, connected space defined by an irregular external potential, 
the definition of the length scale characterizing the decay of superfluid density at the boundaries 
may be generalized by introducing the notion of a ``healing surface.'' For example, in a cylindrically symmetric system
parameterized by height $z$ and radius $0\le R(z) \le R_{t}$ such as we consider in this work (where $R(z)$ is the radial coordinate from the central guiding axis of the cylinder), the local displacement length $d(z)$ in Eq.(\ref{eqn:displength}) 
can be used to define a cylindrically symmetric healing surface. In this case, the healing surface is a surface of revolution defined by the rotating the coordinates $(z,R_{t}-d(z))$ about the $z$-axis.
This healing surface is shown in Fig. \ref{fig:tube_localsf} b) for the potential in Eq.(\ref{eq:tube}) at two different temperatures.
Note that the healing surfaces exhibit fluctuations at the single-atom length scale $\mathcal{O}(1\textrm{ \AA})$ even for well-converged numerical calculations. These fluctuations reflect the numerical error in the radial location of maximal superfluid density at each $z$-coordinate. If $\max_{0\le R \le R_{t}(z)}\rho_{s}(R,z)=0$ for a given $z$, the displacement length is locally undefined at that $z$ coordinate. The local displacement length can be further generalized to quantify the decay of elements of the locally-defined superfluid response tensor $\rho_{s}^{ij}(\vec{r})$ near other types of boundaries and inhomogeneities.

\begin{table}
\caption{\label{table:tube_healing}Global displacement lengths $d$ (\AA) of liquid $^{4}$He subject to the one-body potential 
in Eq.(\ref{eq:tube}) for a uniform cylinder of radius $10$~\AA, calculated from radial local superfluid density according to Eq.(\ref{eqn:displength}). The empirical healing length $\xi(T)$ is shown for comparison. Parameters of the simulations are the same as in Fig.\ref{fig:tube_localsf}. Errors are estimated by calculating the displacement lengths $d_{\pm}$ corresponding to $\rho_{s}(R)\pm \sigma(R)$ in Fig.\ref{fig:tube_localsf}, where $\sigma(R)$ is the local Monte Carlo error.}
\begin{ruledtabular}
\begin{tabular}{@{}lll}
{$T$ (K)}&{$d$ (\AA)}&$\xi(T)$ (\AA) \\ \hline
{}&{} &{}\\
 $0.625$&{3.84}$\substack{+0.02 \\ -0.03}$ &{4.27}\\
 {}&{}&{}\\
 $1.25$&{3.98}$\substack{+0.03 \\ -0.03}$&{6.04}\\
 {}&{}&{}\\
 $1.60$&{4.13}$\substack{+0.03 \\ -0.03}$&{8.33}\\
 {}&{}&{}\\
 $2.00$&{4.27}$\substack{+0.06 \\ -0.05}$&{18.73}\\
\end{tabular}
\end{ruledtabular}
\end{table}
 
In order to define a global displacement length $d$ in a cylindrically symmetric system, the distance $d(z)$ is averaged over $z$. The temperature dependence of the global displacement length for the potential in Eq.(\ref{eq:tube}) is shown in Table \ref{table:tube_healing}. The increase of $d$ with temperature demonstrates the increasing length scale of 
the decay of superfluidity at a nonadsorbing boundary of a nanoscale system. Although data showing the $T\rightarrow T_{\lambda}$ critical scaling of $d$ is lacking, we 
nevertheless observe a smaller rate of increase of $d$ with temperature than is expected from Eq.(\ref{eqn:healing}) for $\xi(T)$.
We note that the displacement length from a planar boundary calculated from the free energy of the Hills-Roberts theory \cite{hillsroberts} also increases more slowly with temperature than the empirical healing length. The displacement length data of Table \ref{table:tube_healing} and Table \ref{table:aperture_healing_table} (see Sec. \ref{sec:apertureresults}) indicate that the length scale characterizing the decay of superfluidity at a boundary of a nanoscale container is less sensitive to the phase transition than the empirical healing length $\xi(T)$, which has a critical exponent that depends on the scaling of the bulk superfluid density as $T\rightarrow T_{\lambda}$. \cite{josephsonscaling} In section \ref{sec:apertureresults}, we show that at a given temperature $T$, $\xi(T)$ is a qualitative lower bound for the radius of a nanoscale aperture connecting two superfluid reservoirs that allows the expected bulk value of the global superfluid fraction to be attained. Therefore, whereas the  theoretical displacement length estimator $d$ quantifies the average displacement of the superfluid from the wall at a given temperature, the empirical healing length quantifies the characteristic radius below which a nanoscale aperture becomes a weak link.

In the next section we describe the results of calculations of superfluid observables for cylinders interrupted by a septum containing one or more nanoscale apertures.  We first
use the estimator of displacement length in Eq.(\ref{eqn:displength}) together with the local and global superfluidity estimators to study the effect of a single nanoaperture interrupting a superfluid reservoir.
Subsequently, we investigate the effect of an array of multiple nanoapertures on the global superfluid fraction of the bipartitioned reservoir system.

\section{\label{sec:apertureresults}PIMC calculations for cylindrical $^{4}$He reservoirs separated by 
nanoscale apertures}

\begin{figure}
\begin{center}
\includegraphics[scale=.56]{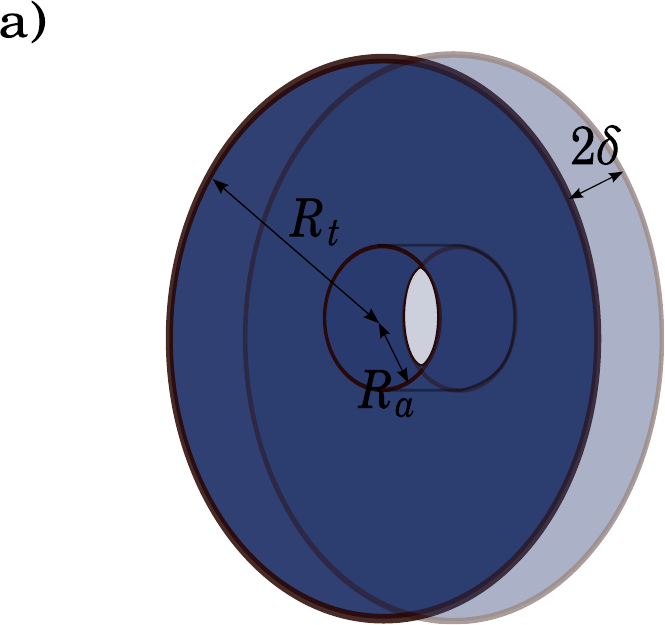}\\
\includegraphics[scale=.50]{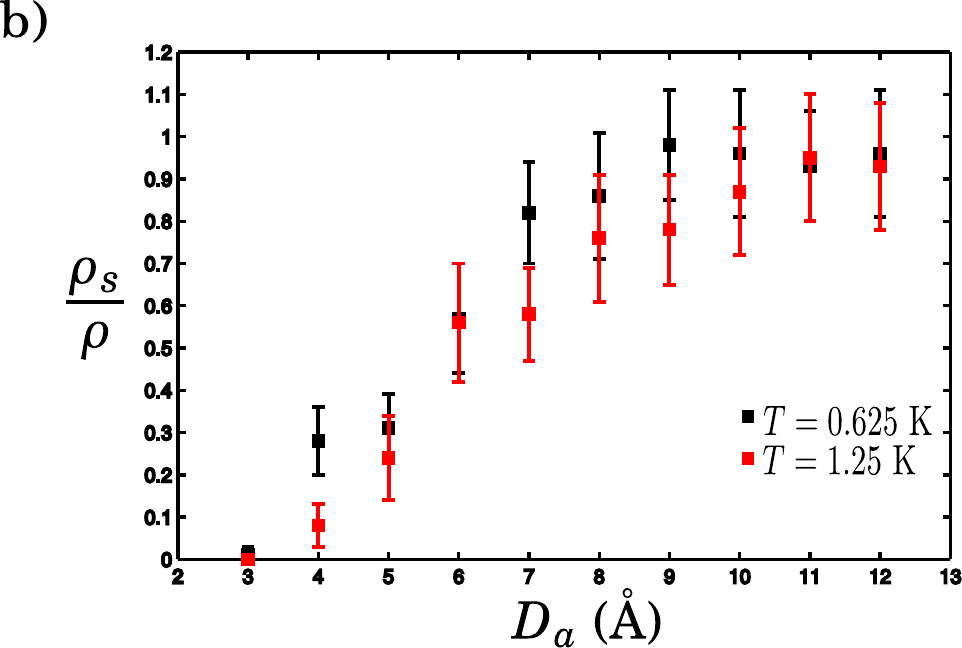}\\
\caption{a) Cross-sectional view of a single aperture in a septum separating cylindrical reservoirs  of liquid $^4$He. $2\delta$ is the thickness of the septum, $R_{t}$ is the radius of the container, $R_{a} = D_{a}/2$ is the radius of the aperture allowing for transport of fluid between the reservoirs.
b) Superfluid fraction of reservoirs ($N=35$, $D_{t}=13$~\AA, 
$L = 24$~\AA) of liquid $^4$He connected by a single atomic-scale to nanoscale aperture
as a function of aperture diameter at temperatures $T=0.625$~K (black squares) and $1.25$~K (red squares).   
In each case, the aperture has thickness $2\delta = 3$~\AA~and all calculations use periodic boundary conditions in $z$.}
\label{fig:aper1}
\end{center}
\end{figure}

An analysis of global superfluid fraction and local superfluid density may be undertaken for reservoirs 
of liquid $^{4}$He separated by a septum pierced with one or more apertures; a cross-sectional view of 
such a septum with a single aperture centered on the cylinder axis is shown in Fig.~\ref{fig:aper1} a). To explore the distribution of superfluidity in reservoirs containing 
such aperture arrays, we consider an external potential given by:
\begin{small}
\begin{eqnarray}
V_{\textrm{{\footnotesize wall}}}^{(N_{a})}(R,\phi, z)&=&\frac{V_0}{2}\Big[1+\tanh\left(\frac{R-R_t}{\sigma_R}\right)\Big] \nonumber \\ &+& 
 \left[ \vphantom{\left(\frac{R_{j}-{D_{a}\over 2}}{\sigma_R}\right)}\frac{V_0}{4}\Big[1+ \tanh\left(\frac{z+\delta}{\sigma_Z}\right)\Big]
\nonumber \right. \\ &\cdot& \left. \Big[1-\tanh\left( \frac{z-\delta}{\sigma_Z} \right) \Big] \nonumber\right. \\     &\cdot& \left. \left( {1\over 2}\right)^{N_{a}}\prod_{j=1}^{N_{a}}\Big[ 1+\tanh\left(\frac{R_{j}-{D_{a}\over 2}}{\sigma_R}\right)\Big]  \right]
\label{eq:aperm}
\end{eqnarray}
\end{small}
which represents a septum of length $2\delta$ located between $z = +\delta$ and $z=-\delta$, pierced by $N_a$ circular apertures of radius $R_{a}=D_{a}/2$ and thickness $2\delta$. The apertures are located at variable positions $R_j, j = 1,...,N_a$,
where $R_{j} = \sqrt{(R\cos\phi -x_{j})^{2}+(R\sin\phi -y_{j})^{2}}$ is the radial distance from the center of the $j$-th aperture
located at coordinates $(x_{j},y_{j})$ to the point of interest.
The remaining potential parameters  
(set equal to the specified fixed values if they are held constant in the simulations) are given as follows:
\begin{enumerate}
\item maximum potential strength: $V_{0}=150$~K
\item tube radius: $R_{t}$  
 \item steepness of potential at cylinder boundary: $\sigma_R = 0.2$~\AA, 
 \item septum thickness: $2\delta =  3.0$~\AA, 
\item steepness of potential at septum boundary: $\sigma_Z = 0.2$~\AA, 
\item number of apertures: $N_{a}$
\item aperture radius: $R_{a} = D_{a}/2$.
\end{enumerate}

In addition, we denote the length of the tube along the $z$-axis by $L$ (periodic boundary conditions are imposed in this direction).  The potential in Eq.(\ref{eq:aperm}) may
in general break the cylindrical symmetry of the system.

We first computed the superfluid fraction in a cylindrical reservoir ($D_{t}=13$~\AA, 
$L = 24$~\AA) of $N=35$ atoms with 
a single intervening aperture of various diameters 
$D_{a}$ (Fig.~\ref{fig:aper1} b)~). In this case, the aperture center coincides with the center of 
the septum (both on the cylinder axis) so that cylindrical symmetry is maintained. At both $T=0.625$~K and $T=1.25$~K, no superflow is observed through apertures 
with diameters less than 3~\AA . As the aperture diameter is increased, the superfluid fraction increases to reach 
the values corresponding to those of reservoirs without a septum (see Fig.~\ref{fig:tube_sf}). At $T=0.625$~K and $T=1.25$~K,
the simulations show non-negligible superflow through the hole with diameter larger than 5~\AA. The empirical healing length of superfluid $^4$He at $T=0.625$~K and $T=1.25$~K is 
calculated from Eq.(\ref{eqn:healing})
to be 4.27~\AA~ 
and 6.04~\AA, respectively. For $D_{a}$ larger than $2\xi(T=0.625~\textrm{ K})$, 
the superfluid fraction reaches unity (within statistical error).
Conversely, $\rho_{s}/\rho$ falls below unity 
for both $T=0.625$~K and $T=1.25$~K when the aperture diameter satisfies $D_{a} \lesssim 2\xi(T)$. Therefore, these data 
are consistent with the 
formation of a superfluid $^{4}$He weak link in a cylindrically symmetric nanoscale channel at a temperature $T$ such that $2\xi(T) \approx D_{a}$. One implication of this result is that in order to decrease the temperature at which an array of nanoscale apertures behaves as a weak link from $T=1.25$~K to $T=0.625$~K, $D_{a}$ must be reduced by $\sim$ 4 \AA, indicating that atomic-scale imperfections in the fabrication of the nanoscale aperture array can affect the sharpness of the critical temperature for weak link formation.

\begin{figure*}
\begin{center}
\includegraphics[scale=.45]{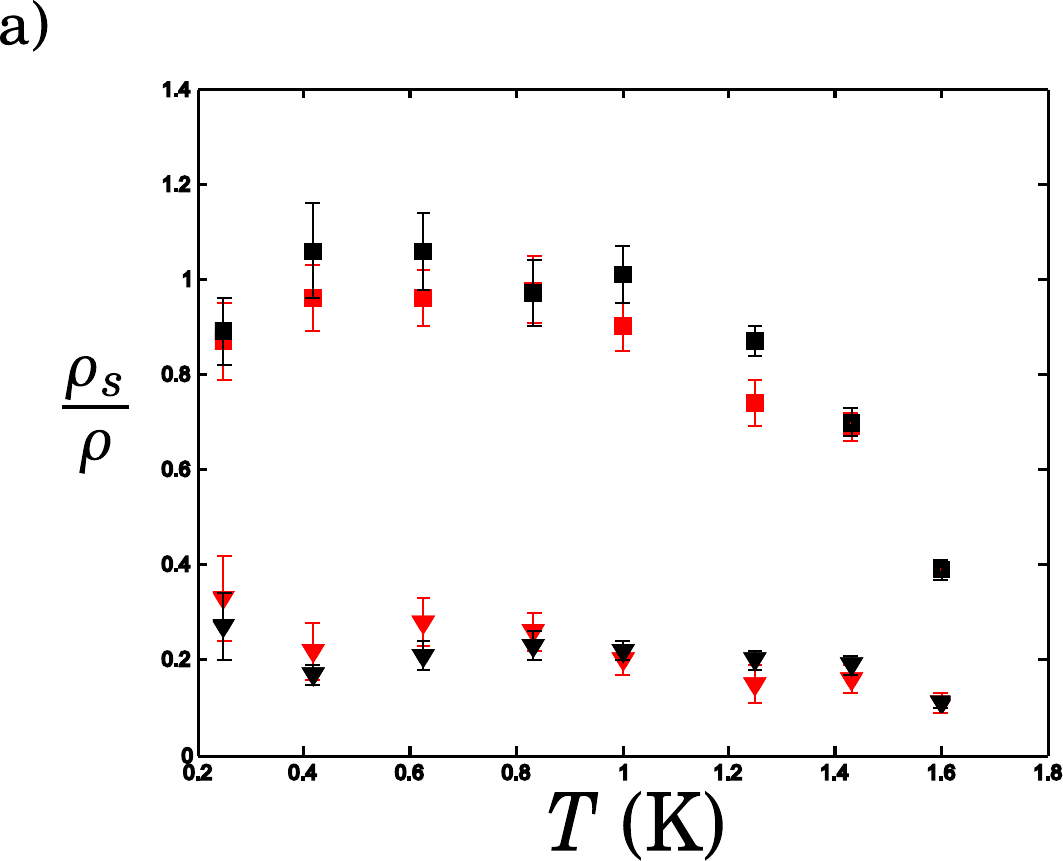}\\
\includegraphics[scale=.35]{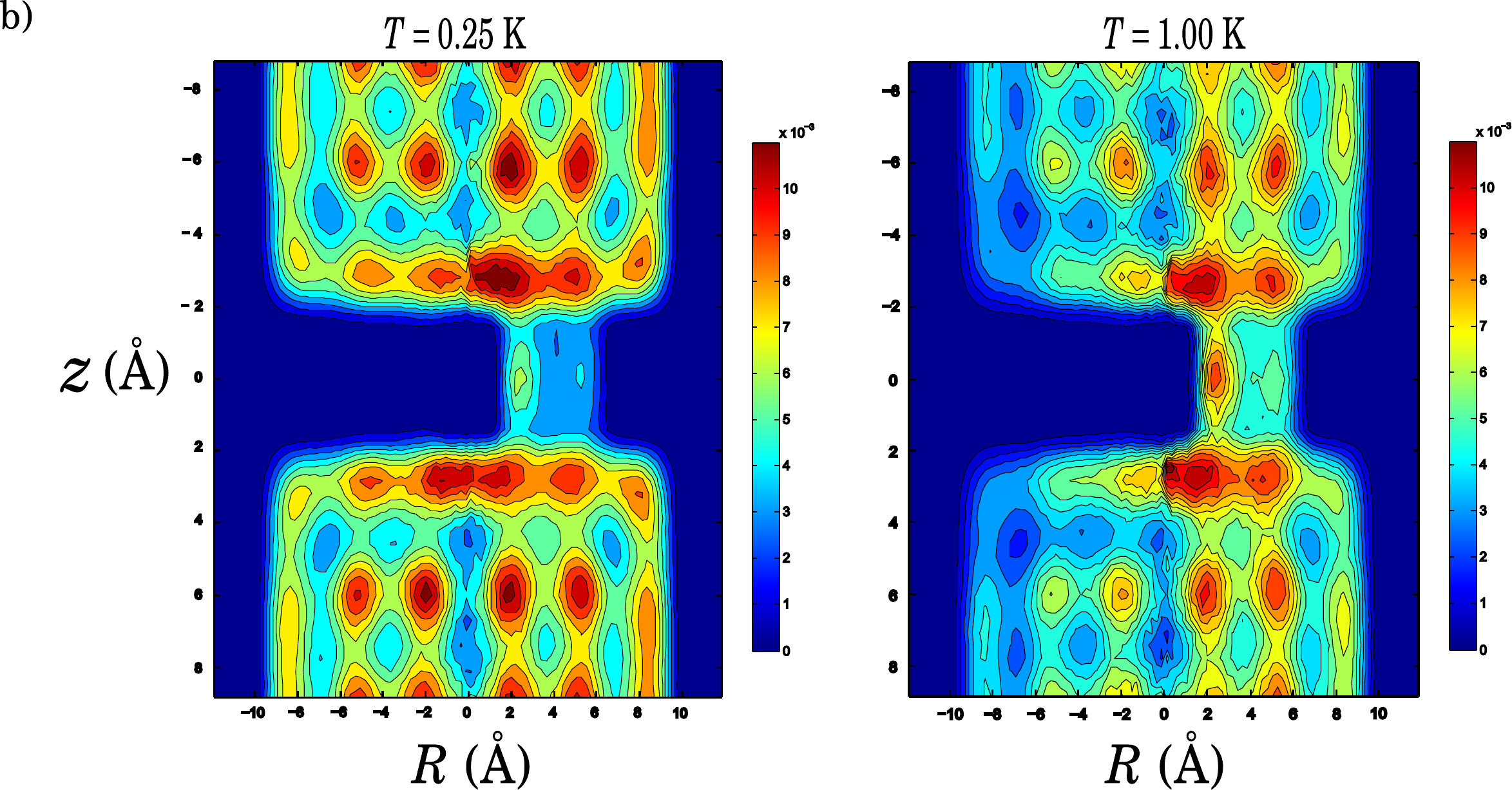}
\caption{ a) Comparison of global transverse superfluid fraction, computed by the projected path area estimator in Eq.~(\ref{eqn:areaest}) (squares), with the longitudinal superfluid fraction, computed by the winding number estimator 
 (inverted triangles), for reservoirs of $^{4}$He ($N=100$, $D_{t}=20$ \AA, $L=18$ \AA) separated by a septum containing a single aperture with $D_{a}=6$ \AA. 
Red symbols denote results with the aperture center on the axis of symmetry; black symbols denote results with the aperture located off the axis of symmetry by 4~\AA.
 b) Local longitudinal superfluid densities 
 $\rho_{s}(z,R)$ (obtained by averaging Eq.(\ref{eq:local_sf}) over the angular coordinate $\phi$), shown as functions of $(z,R)$ in cylinders ($L=18$ \AA, $D_{t} = 20$ \AA)
 with a single off-center aperture having $\delta = 1.5$ and $D_{a}=6$~\AA~at $T=0.25\,$K 
 (left) and 
 $T=1.00\,$K (right).} 
\label{fig:aper2}
\end{center}
\end{figure*}

We proceed to analyze the reduction of the longitudinal superfluid fraction due to the presence of the septum by comparing it to the transverse superfluid fraction, which represents the superfluid response to rotation of the cylinder about its axis of symmetry.
Whereas the longitudinal superfluid fraction is quantified by the winding number estimator in Eq.(\ref{eq:windsf}), the transverse superfluid fraction is written 
in terms of the mean squared projected areas of imaginary-time polymers on the plane perpendicular to the cylinder axis \cite{ceperley95}:
\begin{equation}
\label{eqn:areaest} 
\left( {\rho_{s}\over \rho}\right)_{\perp} = {2mT \langle A_{z}^{2}\rangle \over \lambda I_{c}}. 
\end{equation} 
In Eq.(\ref{eqn:areaest}), $A_{z}$ is the $z$-component of the area vector, having magnitude equal to the area of an imaginary time polymer projected onto the $(R,\phi)$ plane, $\lambda := \hbar^{2}/2m$, and $I_{c}$ is the classical moment of inertia of the polymer. 
Figure \ref{fig:aper2} a) presents a
comparison of the transverse superfluid fraction and the longitudinal superfluid fraction as functions of temperature for a system with a single aperture with radius $D_a = 6$~\AA~and cylinder diameter $D_t = 20$~\AA. The upper traces in Fig. \ref{fig:aper2} a) (square symbols) show $(\rho_{s}/\rho)_{\perp}$, and the lower traces (triangular symbols)  show the $(\rho_{s}/\rho)_{z}$. 
It is evident that, regardless of the aperture location, the transverse superfluid fraction is consistently larger than the longitudinal superfluid fraction and that the former also shows saturation for low $T$, whereas the latter shows only a small increase at lower temperatures and remains less than 0.4 for all temperatures studied. In the present case, this difference reflects the fact that the transverse superfluid flow is not obstructed by any potential that breaks the rotation invariance of the cylinder. In contrast, the longitudinal superfluid fraction is determined by imaginary time paths with nonzero winding number that must pass through the aperture, whatever its location. This constraint severely decreases superfluid response to translations along the cylinder axis.

\begin{table}
\caption{\label{table:aperture_healing_table}
Average displacement lengths $\overline{d}$ (\AA) of liquid $^{4}$He in a bipartitioned reservoir subject to the one-body potential 
in Eq.(\ref{eq:aperm}) with a single aperture, $N_{a}=1$, located on-axis with aperture center at $(0,0,0)$. $\overline{d}$ is calculated as $\overline{d} = (2(L/2 - \delta))^{-1}\left( \int_{-L/2}^{-\delta}dz \, d(z)+ \int_{\delta}^{L/2}dz  \, d(z) \right)$. Parameters of the simulations are the same as for Fig.\ref{fig:aper2} a).}
\begin{ruledtabular}
\begin{tabular}{@{}llllllllll}
{$T$ (K)}&0.250&0.417&0.625&0.833&1.000&1.250&1.430&1.600&2.000\\
$\overline{d}$ (\AA) &{4.39}&{4.62}&{4.78}&{4.95}&{4.98}&{5.34}&{5.42}&{5.93}&{6.38}\\
\end{tabular}
\end{ruledtabular}
\end{table}

To gain insight into this difference, we show 
in Fig. \ref{fig:aper2} a) calculations for two locations of the aperture in the septum (red symbols for the aperture center on the axis of symmetry; black symbols for the aperture center situated 4 \AA~off the axis of symmetry). 
Within statistical error, we see that neither the transverse superfluid fraction (upper traces) nor the longitudinal superfluid fraction (lower traces) is significantly affected by the position of the aperture in the 
septum. However, we observe (Fig.\ref{fig:aper2} b)) that the local distribution of superfluidity in the presence of an off-center aperture is shifted radially relative to the nanotube axis compared to the local distribution of superfluidity
both for an on-axis aperture center and for a cylinder with no intervening aperture array. This asymmetric distribution of superfluidity has the consequence that the liquid $^{4}$He would be expected to exhibit an asymmetric response to, e.g., shear motions of the boundary. In general, given information about the shape of a nanoscale container, one can 
therefore use the calculated local distribution of superfluidity to 
determine the optimal configuration of the aperture array for a particular quantum nanofluidic experiment. Therefore, knowledge of the asymmetry in the local superfluid density is useful for experimental realization and applications of nanoaperture arrays in liquid $^{4}$He.

For a single intervening aperture with a center on the cylinder axis we may apply the displacement length definition of Eq.(\ref{eqn:displength}) to extract an averaged displacement length for the entire bipartitioned reservoir. In Table \ref{table:aperture_healing_table}, we show the cylinder-averaged displacement length 
$\overline{d}$ in a system containing a single aperture with a center on the cylinder axis. 
$\overline{d}$ is calculated by averaging the values of $d_{L}(z)$ and $d_{R}(z)$ obtained by applying Eq.(\ref{eqn:displength}) to regions of the cylinder to the immediate left and right of the septum, respectively.
Similar to the displacement lengths calculated for a system without an aperture array (Table \ref{table:tube_healing}), $\overline{d}$ is observed to increase with temperature. The larger values in Table \ref{table:aperture_healing_table} compared to Table \ref{table:tube_healing} are due to the fact that, for a given tube radius $R_{t}$, the local displacement length $d(z)$ computed for $z$ near a septum is larger on average than $d(z)$ computed for $z$ far from the septum or in the same tube without a septum.

It is clear that the definitions of local and global displacement lengths become ambiguous when the potential is no longer cylindrically-symmetric. For example, in Fig. \ref{fig:aper2} b) the local superfluid density distribution is displaced from the wall with a smaller characteristic length
in the half-cylinder containing the aperture compared to the half-cylinder without it. This result indicates that the superfluid mass density is displaced asymmetrically from the boundaries in a manner which depends 
on the geometry of the confining potential. Global estimates of healing behavior or superfluid density cannot account for the asymmetry; this is why the introduction of the notion of a healing surface, which reveals the local structure of healing, is necessary.

\begin{figure*}
\begin{center}
\includegraphics[scale=.5]{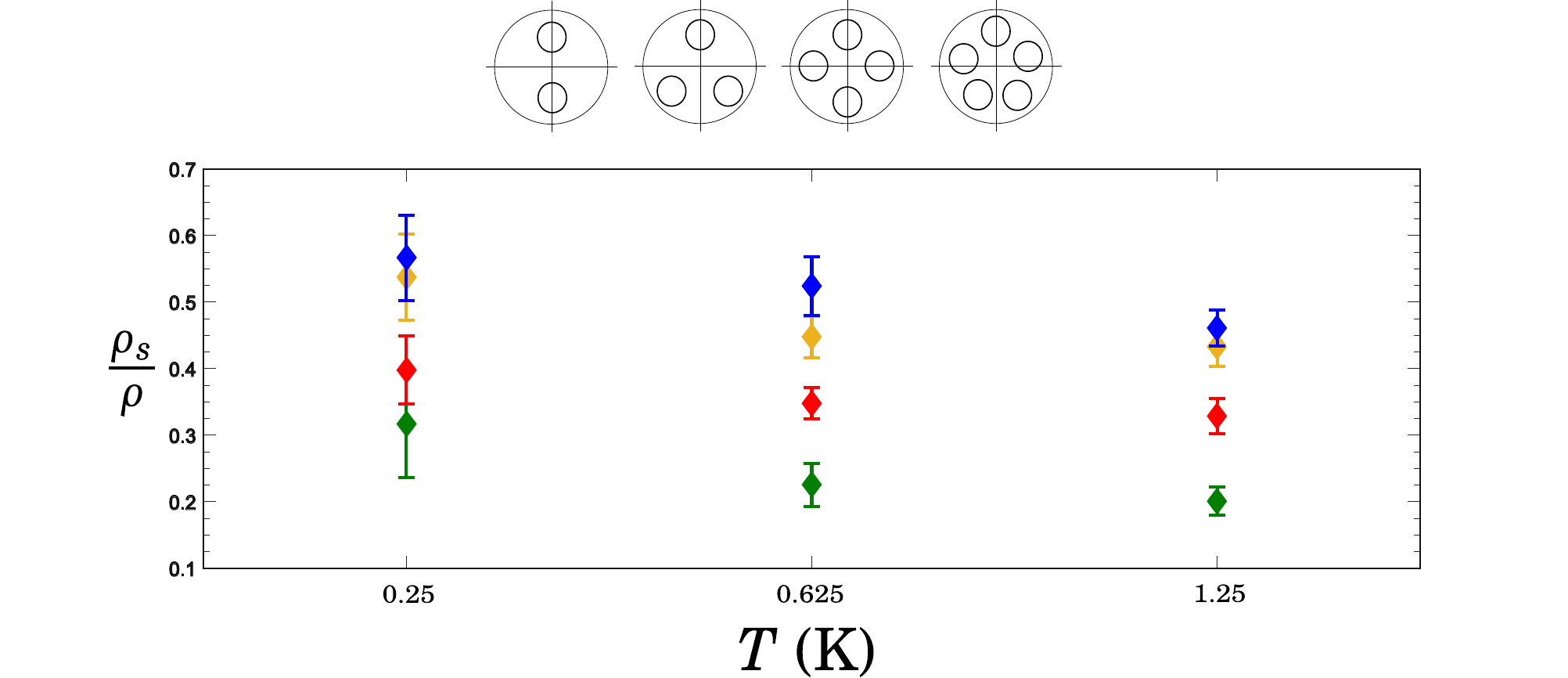}\\
\caption{Global superfluid fractions 
computed by the winding number estimator,  Eq.(\ref{eq:windsf}),
of a reservoir of $N=100$ $^{4}$He atoms bipartitioned by an aperture array defined in Eq.(\ref{eq:aperm}) for $N_{a}=2$ (green), $N_{a}=3$ (red),  $N_{a}=4$ (yellow), $N_{a}=5$ (blue) apertures having $D_{a}=5$ \AA~for three temperatures below $T_{\lambda}$. Configurations of apertures in the septum are shown above the plot. The $(x \textrm{ (\AA)}, y\textrm{ (\AA)})$ coordinates of the aperture centers in the septum plane are: $(0,4),(0,-4)$ for $N_{a}=2$; $(0,5),(4.33,-2.5),(-4.33,-2.5)$ for $N_{a}=3$; $(-5,0),(5,0),(0,5),(0,-5)$ for $N_{a}=4$;   $(5.71,1.86),(0,6),(-5.71,1.86),(-3.53,-4.86),(3.53,-4.86)$ for $N_{a}=5$.  For all simulations, the tube radius and tube length are given by $D_{t}=20$ \AA~and $L=18$ \AA, respectively.}
\label{fig:fig6}
\end{center}
\end{figure*}

Calculations for multi-aperture arrays are more challenging on account of the increased statistical error in all estimators, particularly the local estimators. However the global longitudinal superfluid fraction, Eq.(\ref{eq:windsf}), is sufficiently stable to allow a systematic study with respect to temperature for a range of aperture numbers.  In Figure \ref{fig:fig6}, we now show the global superfluid fraction computed by the winding number estimator for reservoirs separated by arrays containing  
$N_{a}=2,3,4$ or $5$ apertures, with radius $D_a=5$ \AA~ in all cases. The apertures are arranged in each calculation so that the cylindrical symmetry of the reservoir is decreased to $C_{2}$, $C_{3}$, $C_{4}$ and $C_{5}$ symmetry, respectively. 
Several trends are apparent from these results.  First, the global superfluid response is less than the value for the cylinder without any aperture and the superfluid fraction also decreases with temperature, as expected.  Second, it is evident that in general, for a given temperature the superfluid response increases as the number of apertures $N_a$ increases.  It is interesting that this increase with aperture number occurs despite the fact that the aperture radius is smaller than the healing length in these calculations (the empirical healing length values at these temperatures are $\xi(T=0.25) = 3.69$~\AA, $\xi(T=0.625) = 4.1$~\AA, $\xi(T=1.25) = 6.0$~\AA).  Thus the critical factor for the increased longitudinal superfluid response 
with increasing aperture number $N_a$ is that the individual aperture radii  are greater than the helium atomic dimension of length ($D_{a} \gtrsim 4$ \AA), and not whether the helium flow is outside the weak link regime.

\section{\label{sec:conclusion}Summary and Conclusions}

We have used path integral Monte Carlo numerical simulations to analyze the global and local superfluid response of cylindrically-symmetric reservoirs 
of liquid $^{4}$He with and without a bisecting array of nanoscale apertures, using external potentials for the $^{4}$He reservoirs that preclude adsorption at the boundary. 
Global superfluid fractions 
quantifying the superfluid response to translational motion along the cylinder axis (i.e., longitudinal superfluidity)
and rotational motion about the cylinder axis (i.e., transverse superfluidity) were calculated for these systems
by using estimators that are based, respectively, on the longitudinal winding number of imaginary-time polymers representing indistinguishable $^{4}$He atoms, and the projected areas of these imaginary time polymers. We found that the presence of a septum with a single aperture significantly reduces the global longitudinal superfluid response but has a smaller effect on the superfluid response to rotational motion of the aperture, with both of these reductions being approximately independent of the location of the aperture center in the septum. 
Furthermore, the longitudinal superfluid response decreases as the aperture diameter decreases, with a significant drop when the diameter satisfies $2\xi(T) \approx D_{a}$, consistent with the formation of a superfluid $^{4}$He weak link in a cylindrically symmetric nanoscale channel. 
For an aperture array with $N_a >1$ apertures, we found that the longitudinal superfluid fraction increases when the number of apertures $N_{a}$ is increased, regardless of whether individual apertures are operating in the weak link regime or in the lower temperature regime where $2\xi(T) \ll D_{a}$.  
 
We also calculated the local distribution of superfluid density in these nanoaperture array systems by using a positive-definite estimator of local superfluid density (Eq.(\ref{eq:local_sf})) that weights equally all beads participating in imaginary time polymers with nonzero winding number. In contrast to the prediction of Ginzburg-Pitaevskii theory for bulk liquid helium that the superfluid density decreases monotonically near the boundary of a system, we find that, as a consequence of the confining potential, the radial superfluid density does not 
decay monotonically as $R\rightarrow R_{t}$ in a nanoscale cylinder. 
Instead, it shows radial oscillations reflecting the effect of the interatomic interactions.  In systems containing an off-axis intervening aperture, asymmetrical displacement lengths are observed
in the upper and lower halves of the cylinder. This asymmetry could be exploited 
in the design of aperture arrays for experiments in superfluid hydrodynamics.

We analyzed two temperature-dependent length scales related to the global superfluid response and local superfluid density distributions, the empirical temperature-dependent healing length $\xi(T)$ and the theoretical temperature-dependent displacement length $d$. 
Our results indicate that the empirical healing length $\xi(T)$ is qualitatively useful for predicting the temperatures and aperture radius at which superfluidity decreases below the expected bulk value and a weak link can form. However, for a detailed picture of the length scale characterizing the decay of superfluidity at a boundary of a non-translationally invariant system, rather than making empirical estimates, a microscopic estimator is required that can be calculated from the $\rho_{S}(R,z)$ data. The local and 
averaged displacement length estimators given by $d(z)$ and $d$ ($\overline{d}$ in the case of reservoirs separated by a septum), respectively, accurately quantify the decay of superfluidity at a boundary of a cylindrically symmetric container, and they also exhibit an increase with temperature over the range $T = 0.25 - 2.0$~K.   A study of the critical scaling of the displacement length 
$d$ as $T\rightarrow T_{\lambda}$ that takes into account finite-size effects in various confined geometries is an important avenue for future research.

In this work we have considered only static properties of the constrained superfluid. In order to analyze the effects 
of externally imposed flow on the local superfluid density and displacement lengths with the PIMC method, local estimators of velocity and vorticity 
and their correlations must be calculated. We have derived an estimator for the local vorticity in the system 
that will divulge information about the equilibrium structure of line-like defects in superfluid density 
in cylindrically confined systems.\cite{volkoffthesis} Results in this direction will be reported in a future publication.  We expect that the construction of nanoscale aperture arrays similar to those analyzed in this work could lead to experimental observation of Josephson oscillations between phase coherent reservoirs of liquid $^{4}$He deep in the superfluid phase, which may be be exploited in superfluid-based technologies. \cite{packsat}

\begin{acknowledgments}
This work was supported by Basic Science Research Program 
through the National Research Foundation of Korea (NRF) funded by the Ministry of Education (2015R1D1A1A09056745), by the National Science Foundation (NSF) Grant No. PH9-0803429 through the Physics at the Information Frontier Program, and by the National Science Foundation Grant No. CHE-1213141.
TJV acknowledges the NSF/NRF Korea EAPSI program for funding and H. Shin and S. Park for useful suggestions. YK acknowledges the support from the Supercomputing Center/Korea Institute 
of Science and Technology Information with supercomputing resources including technical support 
(KSC-2015-C3-001).
\end{acknowledgments}

\bibliography{yk_tv_kbw_bib}

\end{document}